\documentclass[aps,prl,twocolumn,amsmath,superscriptaddress,floatfix]{revtex4}

\usepackage[dvips]{graphicx}

\newcommand{\micron}{\,\mu\textrm{m}}
\newcommand{\beq}{\begin{equation}}
\newcommand{\eeq}{\end{equation}}
\newcommand{\rd}{\mathrm{d}}

\begin{document}

\title{Liquid Transport Due to Light Scattering}

\author{Robert D. Schroll}
\affiliation{Physics Department and the James Franck Institute, The
University of Chicago, 929 E. 57th St., Chicago, Illinois 60637}
\author{R\'egis Wunenburger}
\author{Alexis Casner}
\altaffiliation[Present address: ]{Commissariat \`a l'Energie
Atomique, BP 12, 91680
Bruy\`eres-le-Ch\^atel, France} \affiliation{Universit\'e Bordeaux I;
CNRS; UMR 5798,  Centre de Physique
Mol\'eculaire Optique et Hertzienne, 351
cours de la Lib\'eration, 33405 Talence Cedex, France}
\author{Wendy W. Zhang}
\affiliation{Physics Department and the James Franck Institute, The
University of Chicago, 929 E. 57th St., Chicago, Illinois 60637}
\author{Jean-Pierre Delville}
\affiliation{Universit\'e Bordeaux I; CNRS; UMR 5798,  Centre de Physique
Mol\'eculaire Optique et Hertzienne, 351 cours de la Lib\'eration, 33405 Talence Cedex, France}

\date{January 5, 2007}

\begin{abstract} 
Using experiments and theory, we show that light scattering by
inhomogeneities in the index of refraction of a fluid can drive a
large-scale flow.  The experiment uses a near-critical,
phase-separated liquid, which experiences large fluctuations in its
index of refraction. A laser beam traversing the liquid 
produces a large-scale deformation of the interface and can cause a
liquid jet to form. We demonstrate that the deformation is produced by a
scattering-induced flow by obtaining good agreements between
the measured deformations and those calculated assuming this
mechanism. 
\end{abstract}


\maketitle

A photon exchanges momentum with its surroundings. Light-scattering
techniques use this effect to probe the structure of materials.  Much
of what we know about the mesoscopic structures in colloidal
suspensions, emulsions, and near-critical fluids have been revealed by
light  scattering~\cite{ackerson75,pine88,vanmegen93,berne00}. More
recently, many researchers have explored how  the intense light beam
generated by a laser can accelerate and trap  micron-sized
particles~\cite{ashkin97,xu04}. Applications range from
laser tweezers~\cite{terray02,grier03,enger04} to particle sorting in
microfluidic 
channels~\cite{macdonald03,ozkan03,neale05}. 
However one consequence of scattering has received little attention.
Since a liquid flows readily, the momentum transferred by light
scattering in a structured fluid can produce a flow along the light
propagation direction. 

In this Letter, we examine one example of a structured fluid, a
phase-separated liquid near a second-order phase transition, and show
that a strong flow  is produced by light-scattering off density
fluctuations in the liquid. This flow is measured indirectly via  the
deformations it produces on the very soft, near-critical liquid
interface. 

\begin{figure}[b!]
\includegraphics{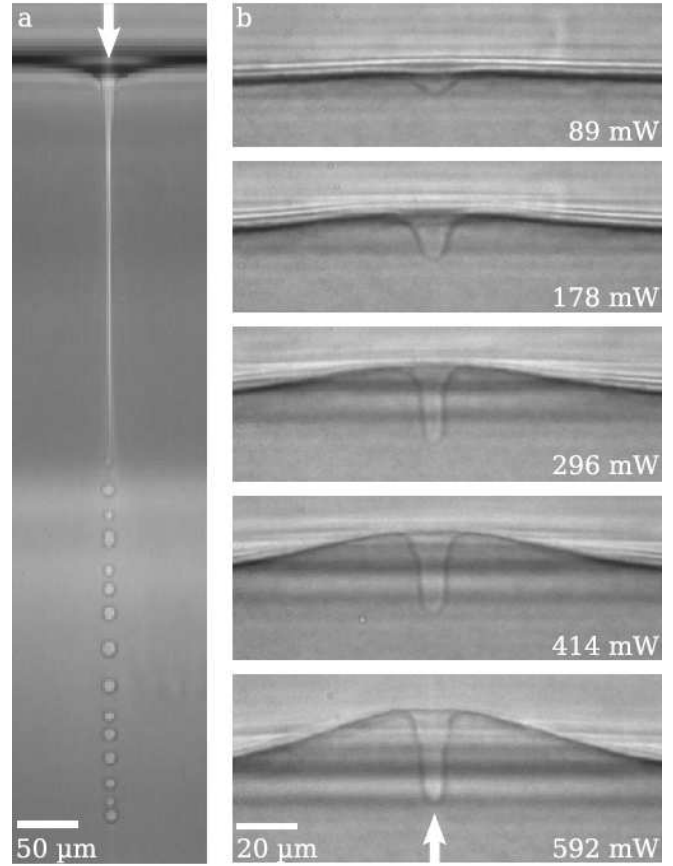} 
\caption{\label{fig:pics} Interface deformations.  (a) When the laser
shines downwards, the upper
fluid (with the larger refractive index) forms a jet intruding into the lower
layer. ($\Delta T = 6\,$K, $\omega_0 =
3.47\micron$, and $P = 490\,$mW) (b) When the laser shines upwards,
the interface forms a downwards tether along the centerline and an upward,
broad hump away from the centerline. ($\Delta T = 1.5\,$K, $\omega_0 =
4.8\micron$, and $P$ as indicated)
The arrows show the direction of propagation of the laser.}
\end{figure}

Figure~\ref{fig:pics} shows the two different types of deformation
observed.  When the laser shines downwards onto the interface, so
that  the beam travels from the phase with the higher refractive index
to the phase with the lower refractive index, a long, thin jet of the
upper-layer liquid forms  along the beam axis and intrudes deep into
the lower fluid  [Fig.~\ref{fig:pics}(a)]~\cite{casner03}.
Sporadically, the end of the jet sheds droplets.  For modest laser
powers, the shedding is regular in time and allows us to measure the
volume flux, which is typically several tens of cubic microns per
second.  When the laser shines upwards, the interface  remains
unbroken even at high power. Instead, a downward tether forms on the
interface due to radiation pressure effects~\cite{casner03b}. Away
from the centerline, the interface also deflects upwards, forming a
hump whose lateral lengthscale is much larger than the beam width
[Fig.~\ref{fig:pics}(b)].  

\begin{figure}
\includegraphics{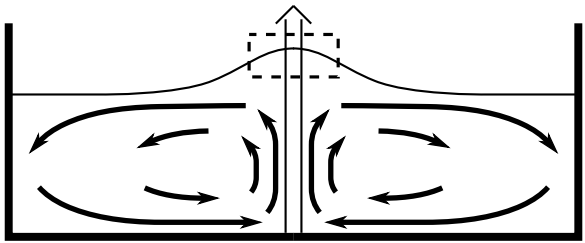} 
\caption{\label{fig:flow} Toroidal recirculation produced in the lower
layer when the laser shines from  below. The hump height is typically around
$10\micron$, while the fluid layer is 1$\,$mm thick. The dotted box indicates the
area seen in the photographs.} 
\end{figure}

Previous works have analyzed how the difference between the refractive
indices of the two liquid phases results in a radiation pressure which
deforms the interface~\cite{casner01,casner03}. While this mechanism
explains the beam-sized deformations, it cannot explain either the 
jet or the broad hump. 
In this Letter, we show that these structures demonstrate the presence
of a bulk flow driven by light scattering.  The rest of this paragraph
gives the key points of our argument, which are also illustrated in
Fig.~\ref{fig:flow}.  We begin with the simpler question of how a
broad hump is created when the interface is illuminated from below. 
Light scattering produces
an upwards body force on the liquid within the laser beam, driving an
upwards flow within the lit region.  By conservation of fluid mass, this flow
is replenished by
a downwards flow.  In the experiment, and in most situations of
interest, inertial effects are negligible. Since purely viscous flows
minimize dissipation~\cite{textbook}, the replenishing flow
takes the form of a single toroidal recirculation.
Viscous stresses associated with the
recirculation deform the interface upwards. This creates a hump whose
width corresponds to the size of the recirculation and is therefore
much wider than the laser beam. 
In the rest of the paper, we show that the qualitative
scenario described above can quantitatively account for the broad hump
shapes measured. In addition, we estimate the volume flux of flow
driven by the light in the jetting regime and find reasonable
agreement with the measured values. 

The fluid used in the experiments is a water-in-oil micellar fluid at
a critical composition.  Above the critical temperature $T_c \approx
35\,^{\circ}\mathrm{C}$, the fluid separates into two immiscible
phases with different micelle volume density $\Phi$, in a second-order
phase transition~\cite{phaseTransition}. Near the critical
temperature, many physical properties scale like power laws in $\Delta
T \equiv T-T_c$.  For our purpose, the most important scaling behavior
is the divergence of the osmotic compressibility, $\chi_T \propto
\Delta T^{-1.24}$. The fluid experiences fluctuations in its order
parameter, $\Phi$, which act as light scatterers.  These fluctuations
have a correlation length $\xi^- \propto \Delta
T^{-0.63}$, which is typically
hundreds of Angstroms. The very weak absorption of light at the laser
frequency ($\alpha_{th} = 3\times 10^{-4}\,$cm$^{-1}$) ensures that
laser heating is negligible. The fluid is enclosed in a thermally
controlled fused quartz cell ($2\times 10\times 40\,$mm$^3$).  Optical
forcing is provided by a linearly polarized TEM$_{00}$ continuous
Ar$^+$ laser, with a vacuum wavelength of $\lambda = 5145\,$\AA\ and a
beam power $P<2\,$W.  The  beam has a Gaussian profile with width
$\omega_0$, varying from $3\micron$ to $15\micron$.  More details
about the experiment can be found in~\cite{casner01}.

\begin{figure}
\includegraphics{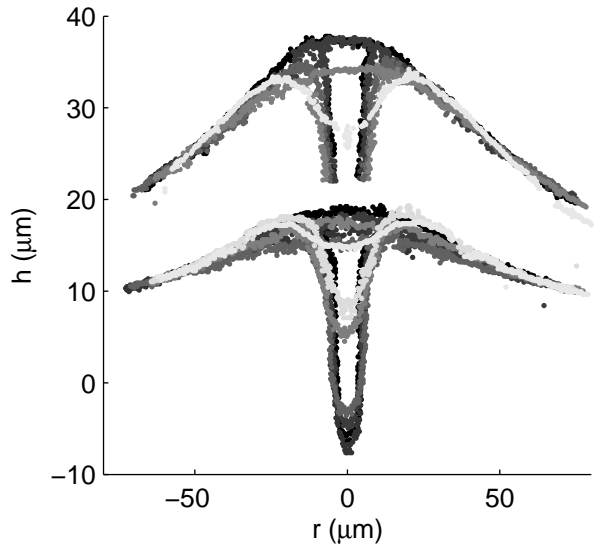} 
\caption{\label{fig:w0} Hump profiles for 
$P=592\,$mW (upper profiles) and for $296\,$mW (lower profiles).
$\omega_0 = 4.8\micron$ (black dots), $5.3\micron$, $6.3\micron$,
$7.5\micron$, $8.9\micron$, $11.7\micron$, and $15.3\micron$
(lightest gray dots) and $\Delta T = 1.5\,$K for all profiles.  
Away from the beam axis, all of the profiles fall onto the same shape,
demonstrating that only the laser power, and not the light intensity,
affects the large-scale hump shape.} 
\end{figure}

We can construct a simple argument for how $u_0$, the strength of the
light-induced flow, depends on the light intensity.  The momentum per
unit volume transferred from the laser beam into the liquid via
scattering is proportional to the beam intensity $I$. Therefore the
body force $F_v$ is also proportional to $I$. This body force acts as
a pressure gradient along the beam axis.
Balancing $F_v$ against viscous resistance
$\mu\, u_0/\omega_0{}^2$, where we use the beam width $\omega_0$ as a 
characteristic lengthscale of the light-induced flow, yields
\beq
u_0 \propto I \omega_0{}^2/\mu \propto P/\mu.
\label{eq:u0}
\eeq
This argument predicts, counter-intuitively,  that the flow strength 
has no dependence on the beam intensity. It only depends  on the beam
power. If this is true, then the interface deformation created by the
flow should also have no dependence on the light intensity.
Figure~\ref{fig:w0} shows two sets of measured hump profiles. Each set
is taken at a fixed beam power, with the different profiles
corresponding to different beam sizes. Remarkably, consistent with the
prediction~(\ref{eq:u0}),  the hump shape away from the center remains
the same for different beam widths when the power is held constant. In
contrast,  the downward-pointing tether at the center of the hump,
previously shown to be created by radiation pressure, varies with the
beam width.

Next, we explicitly calculate the shape of the interface deformation
as a function of $P$ and $\Delta T$ and compare the calculated shapes
with the measured shapes. This requires us to first relate $u_0$, which
characterizes the strength of the recirculation, to the 
laser power $P$, and then to relate the shape deformation to $u_0$.

To obtain $u_0(P)$, we note that 
each scattering off a density fluctuation transfers momentum to the
fluid.  Summing the individual contributions from all fluctuations in
a unit volume gives the body force:
\begin{equation}
F_v = D\,\chi_T\,I \label{eq:force},
\end{equation}
where $D$ describes the interaction of the light with the density
fluctuations and is given by $\frac{\pi^3}{\lambda^4} \frac{n}{c}
\left(\Phi\frac{\partial\epsilon}{\partial\Phi}\right)_T^2 k_B T
\alpha^{-4}$\hspace{0pt}$\times\left[\frac{8}{3}\alpha^3+2\alpha^2+2\alpha
- \left(2\alpha^2+2\alpha +1\right)\ln\left(
1+2\alpha\right)\right]$.  Here, $\alpha \equiv 2(2\pi
n\xi^-/\lambda)^2$ and $n=\sqrt{\epsilon}$ is the index of refraction.
This result comes from  a calculation analogous to the turbidity
calculation by Puglielli and Ford~\cite{ornstein14, puglielli70}.  
Since the correlation length is shorter than the
wavelength of light in the regime examined, the fluctuations act as
Rayleigh scatterers. The calculation also
assumes single scattering, which is justified by the large turbidity
length.
Combining this result with the balance from (\ref{eq:u0}), we
see
\beq u_0 = C\,F_v\,\omega_0{}^2/\mu
	= C\, D \chi_T P/\mu \label{eq:u0C},
\eeq 
where $C$ is an undetermined numerical prefactor dependant on the
details of the coupling of the light-scattering force to the
large-scale flows.  The velocity scale $D \chi_T P/\mu$ is
roughly tens of microns per second for experimental conditions. 

\begin{figure}
\includegraphics{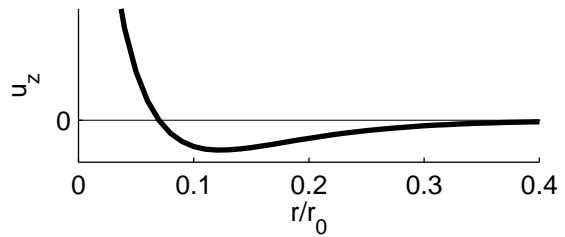}
\caption{Form of the vertical velocity $u_z$ at the mid-plane of the
fluid layer, from the flow model. 
The layer aspect ratio (depth/radius = $L/r_0$) is $1/5$.
\label{fig:uz}} 
\end{figure}

We next relate the flow within the region illuminated by the laser
beam to the large-scale recirculating flow
responsible for the broad hump. The liquid
in the lower layer lies within a cylindrical cell of radius $r_0$ and
depth $L$.  Since the laser beam width is much narrower than the width of the
recirculation, we represent the light-induced flow as simply a point
force along the centerline of a cylindrical cell. 
Also, as the interface
deformations are much smaller than the scale of the flow, we treat the
interface between the layers as perfectly flat. These simplifications
allow us to obtain an analytic solution for the bulk flow.
In our model, the single toroidal recirculation corresponds to an eigenfunction
which satisfies no-slip boundary conditions on the side walls of the
container and free-stress boundary conditions on the top and bottom
surfaces of the container. 
The top and bottom boundary conditions 
are not consistent with the experimental situation. However the error
introduced primarily affects the absolute scale for the strength of the
recirculation, which controls the absolute scale for the height of the
interface deformation. It has little effect on the relative shape of
the deformation or how it changes with the laser power $P$ or the
temperature $T$, which are the aspects we focus on when we compare the
calculation with the measurements.
Figure~\ref{fig:uz} plots the
vertical velocity $u_z(r)$ in the middle of the liquid layer. Note the
flow is upwards at small $r$ and downwards near the side-wall,
consistent with the sketch in Fig.~\ref{fig:flow}.

Finally, we relate the interface deformation $h(r)$ to the viscous
stresses $\sigma_{zz}$ associated with the toroidal
recirculation, and therefore to the laser power $P$.
Because the capillary lengthscale
goes to $0$ near $T_c$,
buoyancy is as  important as surface tension in resisting the deformation.
The steady-state interface $h(r)$ is therefore given by
\beq
2\gamma \kappa(r) + \Delta \rho g h(r) = \sigma_{zz} \equiv C
\,\frac{D \chi_T 
P}{L}\, g(r,r_0/L) 
\label{eq:shape},
\eeq
where $\kappa(r)$ is the mean curvature of the interface. 
To display the various dependencies of
$\sigma_{zz}$, we have rewritten it in terms of a dimensional stress
and two dimensionless quantities, the constant $C$ and the function
$g(r,r_0/L)$ describing 
the radial decay of the stress.
The boundary condition at the wall of
cylindrical cell is $h(r_0) = 0$.  Near the centerline, the interface
develops a downward tether due to radiation pressure. We account for
this $O(\omega_0)$ tether by imposing the boundary  condition $\rd
h/\rd r = 0$ at $r = \omega_0$, which mimics the  observed
circular rim, created by radiation effects which are not included
here.

\begin{figure}
\includegraphics{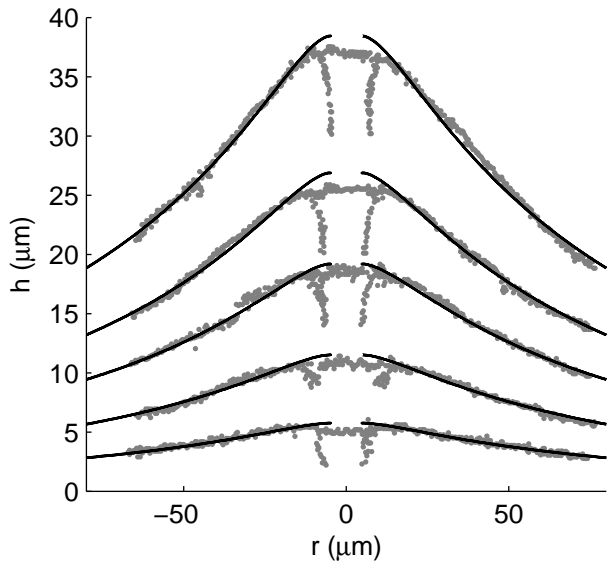} 
\caption{\label{fig:power} The calculated (solid line) and
experimental (dots) hump profile for $\Delta T = 1.5\,$K, $\omega_0
= 4.8\micron$. From bottom to top, $P = {}$88.8, 177.6, 296.0,
414.4, and 592.0~mW.}
\end{figure}

Numerical solutions of (\ref{eq:shape}) at different laser powers are
displayed in Fig.~\ref{fig:power}, together with experimentally
measured interface profiles at the same laser powers. Values of the
material parameters and beam size used in the calculation are taken
from the experiment. The value of the unknown constant $C$ is fixed at
$33$ by requiring that the
maximum height of the  calculated shape equals the measured shape at
$P=592\,$mW.  This is the only fitting
adjustment we have made between the calculation and the measurement.
The agreement between the measured and the calculated interface shapes
is excellent. Comparisons at larger $\Delta T$ for a range of powers 
produced good agreement as well, although the smaller size of the hump
at larger $\Delta T$ makes detailed comparisons 
less precise.  This behavior in $\Delta T$ rules out
thermocapillary flows, used to drive drop motion in previous
studies~\cite{shatz03}, which vanish close to the critical point.

We next consider the liquid transport inside the jet. While the
complexity of the pattern of light propagation at the interface
prevents a detailed comparison between the calculation and the
measurements, it is possible to obtain a rough
estimate. 
Since the jet radius $r_j$ is observed to 
increase weakly with the beam power and is
always less than the beam size $\omega_0$, we assume simply 
that the power of the light trapped inside the jet
is $2 P\, (r_j{}^2/\omega_0{}^2)$, a fraction of the incident beam
power. Given the beam power inside the jet and (\ref{eq:u0C}), the
transport flux is
\begin{equation}
Q = u_0\,\pi\,{r_j}^2 \approx 2\pi\,C\, \frac{D \chi_T P}{\mu} \left(
\frac{r_j{}^4}{\omega_0{}^2} \right). 
\label{qest}
\end{equation}
This estimate gives the right order of magnitude for the volume flux.
For example, an experiment with $P = 473.6\,$mW at $\Delta T = 4\,$K
and $\omega_0 = 5.08\micron$ yields a jet with roughly $1$ $\mu$m
radius and measured volume flux of $110\micron^3/$s. The estimate
(\ref{qest}) gives $310\micron^3/$s. 

In conclusion, we have used a combination of experiment and theory to
demonstrate  that light-scattering can produce a significant flow in a
structured fluid. In the experiment, we measure the large-scale 
interface deformation and the liquid transport produced by
illumination of an intense laser. To show that the deformation is  a
result of light-induced flow, we compare interface deformations
calculated based on the light-scattering mechanism against measured
deformations. Excellent agreements are found between the calculated 
and the measured deformations. We emphasize that 
such light-induced flows exist whenever
fluids have mesoscopic spatial variation in the refractive index and
do not require the fluid to be near a second-order phase
transition. For example, a suspension of $100$~nm-diameter glass beads
in water at $10\%$ volume fraction 
would experience a scattering force $5$ times larger than is seen in our
experiment. While such an effect has been used to transport individual
colloidal particles whose size is comparable with the beam
width~\cite{hart03}, the possibility of transporting smaller colloidal particles
collectively has not been noted before and is worth further
investigation. 

\begin{acknowledgments}
The authors thank B.~Issenmann for providing experimental data on the
jet. This work was supported by a National Science Foundation Graduate
Research Fellowship (RDS), NSF MRSEC DMR-0213745 (U. Chicago), and the Centre
National de la Recherche Scientifique and Conseil R\'{e}gional
d'Aquitaine.
\end{acknowledgments}

\bibliography{jetting}

\end{document}